\newcommand\beq{\begin{equation}}
\newcommand\eeq{\end{equation}}
\newcommand\beqa{\begin{eqnarray}}
\newcommand\eeqa{\end{eqnarray}}
\documentclass{elsart}
   \usepackage{bm}
\usepackage{graphicx,amssymb}
\usepackage{bm}
\usepackage{amsmath}
\journal{Physica A}
\begin{document}
\runauthor{Sabbane, Tij and Santos}
\begin{frontmatter}

\title{Maxwellian gas undergoing a stationary Poiseuille flow in a pipe}
\author{Mohamed Sabbane,}
\author{Mohamed Tij}
\ead{mtij@fsmek.ac.ma}
\address{D\'{e}partement de Physique, Universit\'{e} Moulay Isma\"{\i}l, Mekn\`{e}s,
Morocco}
\author{Andr\'es Santos\corauthref{cor1}}
\ead{andres@unex.es}
\ead[url]{http://www.unex.es/fisteor/andres}
\corauth[cor1]{Corresponding author}
\address{Departamento de F\'{\i}sica, Universidad de Extremadura,
E--06071 Badajoz, Spain}

\date{\today}
\begin{abstract}
The hierarchy of moment equations derived from the nonlinear  Boltzmann equation is solved for a gas of Maxwell molecules undergoing a stationary Poiseuille flow induced by an external force in a  pipe.  The solution is obtained as a perturbation expansion in powers of the force (through third order). A critical comparison is done between the Navier--Stokes theory and the predictions obtained from the Boltzmann equation for the profiles of the hydrodynamic quantities and their fluxes. The  Navier--Stokes description fails to first order and, especially, to second order in the force. Thus, the hydrostatic pressure is not uniform, the temperature profile exhibits a non-monotonic behavior, a longitudinal component of the flux exists in the absence of longitudinal thermal gradient, and normal stress differences are present.
On the other hand, comparison with the  Bhatnagar--Gross--Krook  model kinetic equation shows that the latter is able to capture the correct functional dependence of the fields, although the numerical values of the coefficients are in general  between 0.38 and 1.38 times the Boltzmann values.
A short comparison with the results corresponding to the planar Poiseuille flow is also carried out.
\end{abstract}
\begin{keyword}
Poiseuille flow \sep Non-Newtonian properties \sep Kinetic theory  \sep 
Boltzmann equation \sep Maxwell molecules
\PACS   05.20.Dd \sep 05.60.-k \sep 47.50.+d
\end{keyword}

\end{frontmatter}

\section{Introduction\label{sec1}}       

The Poiseuille flow, first studied  by  Hagen and Poiseuille towards the half of the 19th century,  is still one of the classical examples in fluid dynamics \cite{NOTE:1}. It describes the steady flow in a slab or in a pipe under the action of a longitudinal pressure gradient, what produces a longitudinal macroscopic velocity with a typical parabolic profile in the transverse directions. Essentially the same effect is generated when the longitudinal pressure difference is replaced by a uniform external force  $m \mathbf{g}$ (e.g. gravity) directed longitudinally. This latter mechanism for driving the Poiseuille flow does not produce longitudinal gradients and so is more convenient than the former in computer simulations as well as from  the theoretical point of view, especially to assess  the reliability of the continuum description \cite{KMZ87,ELM94,TTE97,NOTE:2,NOTE:3,NOTE:4,NOTE:6,RC98,NOTE:9,NOTE:5,ATN02,CR01,TE97,ZGA02,NOTE:10}.
The first study of the Poiseuille flow driven by an external force we are aware of was carried out by 
Kadanoff {\em et al.} \cite{KMZ87}, who simulated the flow with the FHP
lattice gas automaton \cite{FHP86} to confirm the validity of a hydrodynamic
description for lattice gas automata.
In the context of a dilute gas,
 Esposito {\em et al.} \cite{ELM94}  studied 
the Boltzmann equation and found that if the force is sufficiently weak there is a solution which converges, in the hydrodynamic limit, to the local equilibrium distribution with parameters given by the stationary solution of the Navier--Stokes (NS) equations.
A generalized Navier--Stokes theory was seen to give a reasonable account of
a fluid composed of molecules with spin when compared with
molecular dynamics simulations \cite{TTE97}.

The first kinetic theory analysis of the Poiseuille flow clearly exhibiting  non-Newtonian behavior was carried out by  Alaoui and Santos \cite{NOTE:2}, who found an exact solution of the Bhatnagar--Gross--Krook (BGK) model kinetic equation for a particular value of $g$. 
 The general solution of the BGK model 
under the form of an expansion in powers of ${g}$ through fifth order was obtained by Tij and Santos
\cite{NOTE:3}. The most interesting outcome of the solution, as first recognized by Malek Mansour {\em et al.} \cite{NOTE:4}, was the presence of a positive quadratic term in the temperature profile to second order in $g$, in addition to the  negative quartic term predicted by the NS description. As a consequence of this new term, 
the temperature
profile exhibits a bimodal shape, namely a local minimum at the middle of
the channel surrounded by two symmetric maxima at a distance of a
few mean free paths. In contrast, the continuum hydrodynamic equations
predict a temperature profile with a (flat) maximum at the middle.
The Fourier law is dramatically violated since in the slab
enclosed by the two maxima the transverse component of the heat flux is
parallel (rather than anti-parallel) to the thermal gradient.
This phenomenon is not an artifact of the BGK model since the same results, except for the numerical values of the coefficients, were derived from an exact solution of the Boltzmann equation for Maxwell molecules through second order \cite{NOTE:6}, as well as from approximate solutions of the Boltzmann equation for hard spheres by Grad's moment method \cite{RC98,NOTE:9}.
It is interesting to note that this correction to the NS temperature profile  is not captured by the next hydrodynamic description, namely the Burnett equations \cite{NOTE:5}. Therefore, in order to describe  a non-monotonic temperature field [which is an $\mathcal{O}(g^2)$-order effect] from a hydrodynamic description one needs to go at least to the super-Burnett equations \cite{NOTE:6}.
A systematic asymptotic analysis of the BGK equation with diffuse boundary conditions to second order in the Knudsen number \cite{ATN02} validates the results of Ref.\ \cite{NOTE:3}  as corresponding to the Taylor series expansion of the normal (or Hilbert) solution around the center of the gap.
 The theoretical predictions of a bimodal temperature profile  have been
confirmed at a qualitative and semi-quantitative level by numerical Monte Carlo 
simulations of the Boltzmann equation \cite{NOTE:4,NOTE:5} and by molecular
dynamics simulations \cite{RC98,CR01}.
In the case of a \textit{dense} gas, even though the non-monotonic behavior of the temperature profile may disappear, the existence of a quadratic term coexisting with the classical quartic term is supported by molecular dynamics simulations \cite{TE97}.
It is worth mentioning that when the Poiseuille flow is driven by a longitudinal pressure gradient rather than by an external force, the NS description is in much better agreement with Monte Carlo simulations of the Boltzmann equation \cite{ZGA02}.

Practically all the studies about the Poiseuille flow driven by an external force have considered the \textit{planar} geometry \cite{KMZ87,ELM94,TTE97,NOTE:2,NOTE:3,NOTE:4,NOTE:6,RC98,NOTE:9,NOTE:5,ATN02,CR01,TE97,ZGA02},  i.e.\ the fluid is assumed to be enclosed between two parallel, infinite plates orthogonal to the $y$-direction, the force acting along the $z$-direction. In the stationary laminar flow the physical quantities have a unidirectional dependence on the $y$ coordinate only. While the planar geometry is the simplest one to study the Poiseuille flow, it is much more realistic to consider that the fluid is enclosed in a cylindrical pipe, the external force being directed along the symmetry axis $z$. In that case, the quantities depend on the distance $r=\sqrt{x^2+y^2}$ from the axis and the fluxes have in general both radial and tangential components.
Two of us have recently solved the BGK model for the cylindrical Poiseuille flow through fourth order in $g$ \cite{NOTE:10}.
The results showed that the structure of the hydrodynamic and flux profiles in the pipe geometry is 
similar to that of the slab geometry. In particular, the temperature
exhibits a non-monotonic behavior as one moves apart from the pipe axis. 
On the other hand, the results are quantitatively sensitive to the geometry of the problem. For instance, the normal stress along the gradient direction is uniform in the planar case ($P_{yy}=\text{const}$), while it has a non-trivial spatial dependence in the cylindrical case ($P_{rr}\neq \text{const}$).
Also, comparison with the BGK solution for the planar  geometry \cite{NOTE:3} shows that  the differences between the kinetic theory results and the NS predictions are in general more pronounced in the planar case than in the cylindrical case.

As is well known, in the BGK model kinetic equation the complicated nonlinear structure of the Boltzmann collision operator is replaced by a single relaxation-time term to the local equilibrium distribution \cite{NOTE:12}. This is usually sufficient to account for many of the nonequilibrium properties of the underlying Boltzmann equation, at least at a \text{qualitative} level \cite{GS03}. However, the existence of only one effective collision frequency does not allow the BGK model to describe  \text{quantitatively} those states where momentum and heat fluxes coexist and are inextricably intertwined, as happens in the Poiseuille flow. This limitation of the BGK model is responsible for an incorrect value $\text{Pr}=1$ of the Prandtl number \cite{NOTE:12},
in contrast to the Boltzmann value  \cite{NOTE:12,NOTE:11} $\text{Pr}\simeq \frac{2}{3}$.
In the planar Poiseuille flow, it is possible to assess the validity of the BGK solution \cite{NOTE:3} by comparing it with the results derived from the Boltzmann equation for Maxwell molecules \cite{NOTE:6} and hard spheres \cite{RC98,NOTE:9}.
The aim of this paper is to fill the gap caused by the absence of results from the Boltzmann equation for the cylindrical Poiseuille flow. We consider a gas of Maxwell molecules and solve the infinite hierarchy of moment equations stemming from the Boltzmann equation through third order in the external force.
The results confirm the correctness of the functional dependence of the hydrodynamic and flux profiles obtained from the BGK model \cite{NOTE:10}. On the other hand, the numerical values of the coefficients are in general different in both kinetic theories, as expected.
The organization of the paper is the following. Section \ref{sec2} is devoted to the description of the flow under study and its solution in a hydrodynamic description to NS order. The kinetic theory description  is presented in Sec.\ \ref{sec3}.
In order to solve the moment hierarchy in a systematic and recursive way, we carry out a perturbation expansion in powers of the external field in Sec.\ \ref{sec4}, the technical details being relegated to Appendix \ref{appB}.
The results are extensively discussed and compared with those of the NS and BGK descriptions in Sec.\ \ref{sec5}. Finally, the main conclusions of the paper are briefly
presented in Sec.\ \ref{sec6}.

\section{Navier--Stokes description of the cylindrical Poiseuille flow\label{sec2}}

The cylindrical Poiseuille flow studied in this paper refers to a monatomic dilute gas enclosed in a long pipe of radius $R$. Let us take the ${z}$-axis as the symmetry axis of the cylinder. The gas particles are subjected to the action of a constant external force per unit mass $\mathbf{g}=g\widehat{\mathbf{z}}$ (e.g. gravity) parallel to ${z}$. After a certain transient stage, the system reaches a steady laminar flow. This nonequilibrium state is characterized by  gradients of the hydrodynamic variables along the directions ${x}$ and ${y}$ orthogonal to the cylinder axis. 
By symmetry, the hydrodynamic fields are expected to depend on $x$ and $y$ through the distance $r=\sqrt{x^2+y^2}$ from the $z$-axis.
In our analysis, we are interested in the bulk region of the flow. This means that the radius $R$ of the cylinder is assumed to be  large enough (as compared with the mean free path) to allow for the existence of such a region.

The \textit{steady-state} balance equations expressing the conservation of momentum and energy are 
\beq
\nabla \cdot \mathsf{P}=\rho \mathbf{g},
\label{new1}
\eeq
\beq
\nabla\cdot \mathbf{q}+\mathsf{P}:\nabla \mathbf{u}=0,
\label{new2}
\eeq
where $\rho $ is the mass density, ${\bf u}=u_z \widehat{\bf z}$ is the flow velocity, ${\sf P}$ is the pressure tensor and ${\bf q}$ is the heat flux vector. 
The geometry of the problem suggests the use of cylindrical coordinates (see Appendix \ref{appA}). Using Eqs.\ (\ref{B5})--(\ref{B9}), Eqs.\ (\ref{new1}) and (\ref{new2}) yield
\beq
\frac{\partial}{\partial r}\left(r P_{rr}\right)=P_{\phi\phi},
\label{2.15a}
\eeq
\beq
\frac{1}{r}\frac{\partial}{\partial r}\left(r P_{rz}\right)=\rho g,
\label{2.15b}
\eeq
\beq
P_{rz} r\frac{\partial
u_z}{\partial r}+\frac{\partial}{\partial r}\left(r q_r\right)=0,
\label{2.16}
\eeq
Equations (\ref{2.15a})--(\ref{2.16}) are exact, but they do not constitute a closed set.

In the Navier--Stokes (NS) description, the momentum and heat fluxes are assumed to be linear functions of the hydrodynamic gradients \cite{NOTE:1,NOTE:12,NOTE:11}. In the geometry of the cylindrical Poiseuille flow the  NS constitutive equations are written as
\beq
P_{rr}=P_{\phi\phi}=P_{zz}=p,
\label{2.20a}
\eeq
\beq
P_{rz}=-\eta\frac{\partial u_{z}}{\partial r},
\label{2.20b}
\eeq
\beq
q_{r}=-\kappa\frac{\partial T}{\partial r},
\label{2.21a}
\eeq
\beq
q_z=0.
\label{2.21b}
\eeq
In Eq.\ (\ref{2.20a}), $p=\frac{1}{3}\text{tr}\,\mathsf{P}$ is the hydrostatic pressure. Equation (\ref{2.20b}) is Newton's friction law, $\eta$ being the shear viscosity. In addition,  Eq.\ (\ref{2.21a}) is Fourier's law, where $\kappa$ is the thermal conductivity and $T$ is the temperature. The latter is related to the number density $n=\rho/m$ (where $m$ is the mass of a particle)  and the pressure $p$ by the (local) equilibrium equation of state. In particular, for a dilute gas, $p=nk_BT$, $k_B$ being the Boltzmann constant. Equation (\ref{2.20a}) implies the absence of normal stress differences in a sheared Newtonian fluid, while Eq.\ (\ref{2.21b}) means that the heat flux is parallel to the thermal gradient in a fluid obeying Fourier's law.
The combination of Eqs.\ (\ref{2.15a})--(\ref{2.21a}) gives the following closed set of hydrodynamic equations:
\beq
\frac{\partial p}{\partial r}=0,
\label{2.22}
\eeq
\beq
r^{-1}\frac{\partial}{\partial r}\left(r\eta\frac{\partial u_z}{\partial
r}\right)=-\rho g,
\label{2.23}
\eeq
\beq
\frac{\partial}{\partial r}\left(r\kappa\frac{\partial T}{\partial
r}\right)=-\eta r \left(\frac{\partial u_z}{\partial
r}\right)^2.
\label{2.24}
\eeq
In principle, the NS hydrodynamic equations (\ref{2.22})--(\ref{2.24}) are too complicated to obtain its explicit solution for arbitrary $g$.
On the other hand, the solution can be found as a
series expansion in powers of $g$. To third order the result is
\beq
p(r)=p_0,
\label{2.24.1}
\eeq
\beq
u_z(r)=u_0-\frac{\rho_0
g}{4\eta_0}{r}^2\left(1+\frac{1}{144}\frac{\rho_0^2g^2}{\eta_0\kappa_0T_0}
r^4\right)+\mathcal{O}(g^5),
\label{2.25}
\eeq
\beq
T(r)=T_0-\frac{1}{64}\frac{\rho_0^2 g^2}{\eta_0\kappa_0}{r}^4
+\mathcal{O}(g^4),
\label{2.26}
\eeq
where we have particularized to a dilute gas of Maxwell molecules, in which case $\eta\propto T$ and $\kappa\propto T$ \cite{NOTE:12,NOTE:11}.
The
subscript 0 in Eqs.\ (\ref{2.24.1})--(\ref{2.26})  denotes quantities evaluated at $r=0$, i.e. $u_0=u_z(0)$, $T_0=T(0)$, $\eta_0=\eta(T_0)$, \ldots.
The corresponding expressions for the fluxes are
\beq
P_{rz}(r)=\frac{\rho_0 g}{2}{r}\left(1+\frac{1}{192}\frac{\rho_0^2
g^2}{\eta_0\kappa_0T_0}{r}^4\right)+\mathcal{O}(g^5),
\label{2.26.1}
\eeq
\beq
q_{r}(r)=\frac{1}{16}\frac{\rho_0^2 g^2}{\eta_0}{r}^3+\mathcal{O}(g^4),
\label{2.26.2}
\eeq
complemented with Eqs.\ (\ref{2.20a}) and (\ref{2.21b}). It is interesting to note that  the spatial variable $r$ can be eliminated between Eqs.\
(\ref{2.25}) and (\ref{2.26}) to obtain the following
nonequilibrium ``equation of state'':
\beq
T=T_0-\frac{\eta_0}{4\kappa_0}(u_z-u_0)^2+\mathcal{O}(g^4).
\label{2.27}
\eeq
By equation of state we mean in this context a relationship holding locally
among the hydrodynamic fields ($p$, $u_z$ and $T$) and that does not contain $g$ explicitly (at least up to a certain order). Since the pressure is uniform
in the NS description, it does not enter into Eq.\ (\ref{2.27}).
Equation (\ref{2.27}) shows that in the Poiseuille flow the velocity and temperature profiles are closely related.

{}From Eqs.\ (\ref{2.24.1})--(\ref{2.26}) we can obtain the mass rate of flow in the NS description. Let us consider a circular section of radius $a$ centered on the cylinder axis and orthogonal to it. The mass of fluid flowing across this surface per unit time is defined, in the reference frame moving with the fluid at $r=a$, as
\beq
\dot{M}(a)=2\pi \int_0^a dr\, r\rho(r)\left[u_z(r)-u_z(a)\right].
\label{new4}
\eeq
Using Eqs.\ (\ref{2.24.1})--(\ref{2.26}) and taking into account that $\rho=mp/k_BT$, one gets
\beq
\dot{M}(a)=\frac{\pi \rho_0^2 g a^4}{8\eta_0}\left(1+\frac{5}{384}\frac{\rho_0^2g^2}{\eta_0\kappa_0T_0}
a^4\right)+\mathcal{O}(g^5).
\label{new5}
\eeq

Before closing this Section, let us make a few remarks about our use of the term ``Navier--Stokes solution''. By that we mean the solution to the  coupled set of hydrodynamic equations (\ref{2.22})--(\ref{2.24}). In fluid mechanics, on the other hand, it is usual to adopt a more restrictive point of view in which by Navier--Stokes equation one means the momentum equation  with a \textit{constant} viscosity $\eta$ and a constant density $\rho$ (``incompressible fluid''). In that case, the solution of Eq.\ (\ref{2.23}) would simply be
\beq
u_z(r)=u_0-\frac{\rho
g}{4\eta}{r}^2,
\label{new3}
\eeq
regardless of the value of $g$. Accordingly,  the mass rate of flow would be
\beq
\dot{M}(a)=\frac{\pi \rho^2 g a^4}{8\eta}.
\label{new6}
\eeq
Equation (\ref{new3}) gives the typical parabolic velocity profile under Poiseuille flow, while Eq.\ (\ref{new6}) is known as Poiseuille's law \cite{NOTE:1}. 
Nevertheless, the shear viscosity $\eta$ depends in general on the temperature $T$ and so the assumption of a constant viscosity implicitly implies a constant temperature.  The same conclusion follows from Eq.\ (\ref{2.24.1}) and the incompressibility assumption $\rho=\text{const}$. The energy equation (\ref{2.24}), however, shows that a velocity profile is compatible with a constant temperature only if the thermal conductivity $\kappa$ is allowed to be infinite. As a matter of fact, Eqs.\ (\ref{2.25}) and (\ref{new5}) reduce to Eqs.\ (\ref{new3}) and (\ref{new6}), respectively, if we formally set $\kappa_0\to\infty$. Since in a dilute gas the relative strength of the thermal conductivity is comparable to that of the shear viscosity (as measured by the Prandtl number $\text{Pr}\equiv 5k_B\eta/2m\kappa\simeq \frac{2}{3}$), the energy equation (\ref{2.24}) cannot be neglected and hence $T\neq \text{const}$.

In summary, in the remainder of this paper we will refer to Eqs.\ (\ref{2.22})--(\ref{2.24}) as Navier--Stokes equations and will compare their solution for a dilute gas of Maxwell molecules [Eqs.\ (\ref{2.24.1})--(\ref{2.26})], along with the momentum and energy fluxes [Eqs.\ (\ref{2.20a}), \ref{2.21b}), (\ref{2.26.1}) and (\ref{2.26.2})], with the results derived from the Boltzmann equation.

\section{Kinetic theory description \label{sec3}}

In kinetic theory,  the relevant information about the nonequilibrium state of a dilute gas is contained in the one-body distribution function  $f({\bf r}, {\bf v}; t)$ \cite{NOTE:11}. Its temporal evolution is governed by the non-linear Boltzmann equation \cite{NOTE:12,NOTE:11}, which in standard notation reads
\begin{eqnarray}
\label{ab8}
\frac{\partial}{\partial t}f+{\bf v}\cdot \nabla f+
{\bf g}\cdot\frac{\partial}{\partial{\bf v}}f&=&
\int d{\bf v}_1 \int d\Omega\, |{\bf v}-{\bf v}_1|
\sigma(|{\bf v}-{\bf v}_1|,  \theta)[f'f_1'-ff_1]\nonumber\\
 & \equiv & J[f,f].
\end{eqnarray}
The influence of the interaction potential  appears through the
dependence of the differential cross section $\sigma$ on the relative velocity
$|{\bf v}-{\bf v}_1|$ and the scattering angle $\theta$. In particular, for Maxwell molecules $\sigma(|{\bf v}-{\bf v}_1|,  \theta)\propto |{\bf v}-{\bf v}_1|^{-1}$.

The knowledge of $f$ in a dilute gas is sufficient to determine its macroscopic state. For example, the hydrodynamic variables and their fluxes are just velocity moments of the distribution function. Thus, the  local number density $n$  is defined by
\begin{equation}
 n = { \int } d {\bf v} \, f .
\label{16}
\end{equation}
Analogously, the flow velocity $\bf u$ and the hydrostatic  pressure $p$ are given by the following expressions:
\beq
{\bf u} =\frac{1}{n} { \int } d {\bf v\, v} f, 
\label{17}
\eeq
\beq
p =\frac {m} {3} { \int } d{\bf v}\, V^{2} f,
\label{18}
\eeq
where the  peculiar velocity $ {\bf V=v-u}$ has been  introduced as the velocity of a particle relative to the flow velocity. The  momentum and energy fluxes are given by the pressure tensor $ {\sf P} $ and the  heat flux vector ${\bf q}$, respectively,  as follows:
\beq
{\sf P} = m { \int } d{\bf v \, V } {\bf V} f, 
\label{19}
\eeq
\beq
{\bf q} = \frac {m} {2}  { \int } d{\bf v}\, V^2 {\bf V} f.
\label{20}
\eeq

In the case of the stationary Poiseuille flow in a pipe, the Boltzmann equation (\ref{ab8}) becomes
\begin{equation}
v_x \frac{\partial f}{\partial x} + v_y \frac{\partial f}{\partial y} + g \frac{\partial f}{\partial v_z} = J \lbrack f, f \rbrack .
\label{22}
\end{equation}
This equation is invariant under the transformations
\beq
(x,v_x)\longleftrightarrow (-x,-v_x),\quad
(y,v_y)\longleftrightarrow (-y,-v_y),\quad (v_z,g)\longleftrightarrow (-v_z,-g).
\label{n2c}
\eeq
Strictly speaking, Eq.\ (\ref{22}) must be supplemented by the appropriate boundary conditions at $r=R$ describing the interaction of the particles with the cylinder surface. However, in this work we are interested in the \textit{bulk} region of the system ($0<r<R-\delta$, where $\delta$ is the width of the boundary layer), which is not affected by the details of the boundary conditions. This implies that the radius $R$ of the cylinder must be sufficiently large, compared to the mean free path of a particle, to allow the existence of such a region (small Knudsen number). 
Here we will assume that this is the case and will look for the Hilbert-class or \textit{normal} solution to Eq.\ (\ref{22}), namely a solution where all the spatial dependence of the distribution function takes place through a \textit{functional} dependence of $f$ on the hydrodynamic fields $n$, $\mathbf{u}$ and $T$.

Since the hydrodynamic variables of the gas and the associated fluxes are the first few moments of the function $f$, it is convenient to consider the hierarchy of moment equations stemming from the Boltzmann equation (\ref{22}). A moment of an arbitrary degree $k=k_1+k_2+k_3$ is defined as
\begin{equation}
  M_{k_1,k_2,k_3} (x,y)= { \int } d{\bf v}\, V^{k_1}_{x} V^{k_2}_{y} V^{k_3}_{z} f(x,y,{\bf v}).
\label{23}
\end{equation}
The invariance properties (\ref{n2c}) imply the parity relations
\beqa
 M_{k_1,k_2,k_3} (x,y;g)&=&
(-1)^{k_1}M_{k_1,k_2,k_3} (-x,y;g)\nonumber\\
&=&(-1)^{k_2}M_{k_1,k_2,k_3} (x,-y;g)
\nonumber\\
&=&(-1)^{k_3}M_{k_1,k_2,k_3} (x,y;-g).
\label{n3}
\eeqa
Although the problem at hand calls for the use of cylindrical coordinates, we will use Cartesian coordinates to construct the moment hierarchy. The reason is two-fold. First, the collision integrals are easier to express in Cartesian (or even spherical) coordinates than in cylindrical coordinates. Second, the use of Cartesian coordinates will serve us as a test of the calculations since, when obtaining the cylindrical components according to Eqs.\ (\ref{2.17}) and (\ref{2.18}), the relevant quantities must depend on $x$ and $y$ through the variable $r=\sqrt{x^2+y^2}$.

The Boltzmann hierarchy of moments  is obtained by multiplying both sides of  Eq.\ (\ref{22}) by $V^{k_1}_{x} V^{k_2}_{y} V^{k_3}_{z}$ and  then integrating over the velocity space. The resulting expression is
\beqa
  \frac{\partial}{\partial x} M_{k_1+1,k_2,k_3} + \frac{\partial}{\partial y} M_{k_1,k_2+1,k_3} 
&+& k_3 \frac{\partial u_z}{\partial x} M_{k_1+1,k_2,k_3-1}
+ k_3 \frac{\partial u_z}{\partial y} M_{k_1,k_2+1,k_3-1}\nonumber\\
& -& g k_3 M_{k_1,k_2,k_3-1} = J_{k_1,k_2,k_3},
\label{24}
\eeqa
where
\begin{equation}
\label{b4}
J_{k_1,k_2,k_3}=\int d{\bf v}\, V_x^{k_1}V_y^{k_2}V_z^{k_3}J[f,f].
\end{equation}
In the sequel, we will use the roman boldface ${\bf k}$ to denote the triad    
$\{k_1,k_2,k_3\}$ and the italic lightface $k$ to denote the sum 
$k_1+k_2+k_3$.
Thus, $M_{\bf k}\equiv M_{k_1,k_2,k_3}$ is a moment of degree $k\equiv k_1+k_2+k_3$.

The collision integral $J_{\bf k}$ involves in general all the velocity moments of the distribution (including those of a degree higher than $k$) and its explicit expression in terms of those moments is unknown. 
An important exception is provided by the Maxwell interaction potential $\varphi(r)=K/r^4$. In that case,
the collision rate is independent of
the velocity, i.e. $|{\bf v}-{\bf v}_1|\sigma(|{\bf v}-{\bf v}_1|,  
\theta)=
\sigma_0(  \theta)$, and thus
$J_{\bf k}$
can be expressed as a bilinear combination of moments of degree equal to or 
smaller than
$k$ \cite{NOTE:7,McLennan}:
\begin{equation}
\label{b5}
J_{\bf k}={\sum_{{\bf k'},{\bf k''}}}^\dagger C_{{\bf k'},{\bf k''}}^{\bf k}
M_{\bf k'}M_{\bf k''},
\end{equation}
where the dagger denotes the constraint $k'+k''=k$.
The coefficients $C_{{\bf k'},{\bf k''}}^{\bf k}$ are linear combinations of
the eigenvalues \cite{McLennan,Alterman}
\begin{equation}
\label{b8}
\lambda_{r\ell}=\int d\Omega\, \sigma_0(  \theta)
\left[1+\delta_{r0}\delta_{\ell 0}
-\cos^{2r+\ell}\frac{\theta}{2}P_\ell\left(\cos\frac{\theta}{2}\right)
-\sin^{2r+\ell}\frac{\theta}{2}P_\ell\left(\sin\frac{\theta}{2}\right)\right]
\end{equation}
of the linearized collision operator, where $P_\ell(x)$ are Legendre polynomials.
The thermal conductivity and shear viscosity for Maxwell 
molecules (first obtained by Maxwell himself) 
are \cite{NOTE:11}
\begin{equation}
\label{b6}
\kappa(T)=\frac{5k_B}{2m}\frac{p}{n\lambda_{11}},
\end{equation}
\begin{equation}
\label{b7}
\eta(T)=\frac{p}{n\lambda_{02}},
\end{equation}
where $\lambda_{02}=\frac{3}{2}\lambda_{11}=0.436\times 3\pi\sqrt{2K/m}$.
The eigenvalues $\lambda_{r\ell}$ can also be expressed as linear combinations of the integrals \cite{NOTE:7}
\beq
A_{2s}=\int d\Omega\, \sigma_0(  \theta)\sin^{2s}\frac{\theta}{2}\cos^{2s}\frac{\theta}{2}.
\label{n1}
\eeq
In particular, $\lambda_{02}=3A_2$, $\lambda_{04}=7A_2-\frac{35}{4}A_4$ and $\lambda_{06}=\frac{27}{2}A_2-\frac{357}{8}A_4+\frac{231}{8}A_6$. 
The first few ratios $A_{2s}/A_2$ are 
 $A_4/A_2=0.15778\ldots$, $A_6/A_2=0.031196\ldots$ and $A_8/A_2=0.0066540\ldots$.
The general expression (\ref{b5}) for $J_{\bf k}$ was elaborated by Truesdell and Muncaster \cite{NOTE:7},  who also give $J_{\mathbf{k}}$ up to $k=5$. A simplification algorithm for this formula, along with explicit expressions of  $J_{\mathbf{k}}$ for $k=6$--$8$, has been recently proposed \cite{NOTE:8}.

 To simplify the subsequent analysis, it is suitable to use dimensionless quantities. 
First, let us define an \textit{effective} collision frequency as
\begin{equation}
\label{b9}
\nu=n\lambda_{02}.
\end{equation}
Next, without loss of generality, we choose a
 reference frame stationary with the flow at $r=0$, i.e. $u_0=0$. Now we introduce the dimensionless variables
\beqa
&& g^*= v^{-1}_0 \nu^{-1}_0 g, \quad
u^*_z=v^{-1}_0 u_z,\quad
p^*=p^{-1}_0 p, \quad
T^*=T^{-1}_0 T, \nonumber\\
&&M^{*}_{\bf k}=n^{-1}_0 v^{-k}_0 M_{\bf k},\quad
J^*_{\bf k}=n^{-1}_0 v^{-k}_0\nu^{-1}_0 J_{\bf k}, \quad
\mathbf{r}^*=\nu_0 v_0^{-1} \mathbf{r}. 
\label{37}
\eeqa
In the above expressions, $ v_0=(k_B T_0/m)^{1/2}$ is the thermal velocity   and, as in Eqs.\ (\ref{2.24.1})--(\ref{2.27}), the subscript $0$ means that the quantities are evaluated at $r=0$. The reduced (gravity) acceleration $g^*$ has a clear physical meaning.
The quantity
$h_0\equiv
v_0^2/g$ is the so-called scale height \cite{textbooks}, i.e.\ the
characteristic distance associated with the external (gravity) field.
Thus, $g^*=(v_0/\nu_0)/h_0$ represents the ratio between the mean free path and the  distance through which a typical particle undergoes the action of ${\bf g}$. While the parameter $g^*$  is a measure of the field strength on the 
scale of the mean free path, the Froude number
\beq
\text{Fr}=\frac{v_0^2}{gR}=\frac{h_0}{R}
\label{new8}
\eeq
measures the strength 
on the scale of the radius of the pipe. The relationship between both dimensionless parameters is
\beq
\text{Fr}=\frac{\text{Kn}}{g^*},
\label{new9}
\eeq
where $\text{Kn}\equiv (v_0/\nu_0)/R$ is the Knudsen number.

The solution of the hierarchy  of equations (\ref{24}) is a very difficult task, even for Maxwell molecules, due to its nonlinear character and the fact that moments of degree $k$ are coupled to the first spatial derivatives of moments of degree $k+1$.  On the other hand, we will restrict ourselves to  cases where the parameter $g^*$ is weak. This assumption is well justified in most of the practical situations, where values $g^*>10^{-2}$ would be abnormally large.  For instance, in the case of air at room temperature under the action of terrestrial gravity, $g^*\sim 10^{-12}$. 
Consequently, we can solve the hierarchy  (\ref{24}) by means of a perturbation expansion in powers of $g^*$, in a similar way as done in other works \cite{NOTE:3,NOTE:6,NOTE:10,NOTE:13}.

\section{Expansion in powers of the external force\label{sec4}}
For the sake of clarity, the asterisks will be dropped in this Section,
so all the quantities will be understood to be expressed in reduced units [cf.\ Eqs.\ (\ref{37})],
unless stated otherwise.

Before performing the  power series expansion of the moments, it is convenient to take into account the symmetry properties (\ref{n3}).   For example, it is obvious that the profiles of the temperature and the hydrostatic pressure are even functions of $g$, while the flow velocity $u_z$ is an odd function. Therefore, we can write
\beq
p=1+p^{(2)} g^2+p^{(4)} g^4+\cdots,
\label{40}
\eeq
\beq
 T=1+T^{(2)} g^2+T^{(4)} g^4+\cdots,
\label{41}
\eeq
\beq
 u_z=u^{(1)} g+u^{(3)} g^3+\cdots.
\label{42}
\eeq
The expansion of the number density $n=M_{000}=p/T$ can be obtained from Eqs.\ (\ref{40}) and (\ref{41}).
These hydrodynamic fields are (even) functions of $x$ and $y$ through the distance $r=\sqrt{x^2+y^2}$.
As for the moments $M_{\bf k}$, they are even (odd) functions of $g$ if their index $k_3$ is even (odd). Thus,
\beq
 M_{\bf k}= M^{(0)}_{\bf k}+M^{(1)}_{\bf k}g +M^{(2)}_{\bf k}g^2+M^{(3)}_{\bf k}g^3+\cdots,
\label{39}
\eeq
where 
$M^{(s)}_{\bf k}=0$ if $s+k_3=\text{odd}$. The zeroth-order term  $M^{(0)}_{\bf k}$ is the moment of the equilibrium distribution function normalized to $p^{(0)}=1$, $T^{(0)}=1$. Its expression is
\begin{equation}
M^{(0)}_{\bf k} =C_{k_1}C_{k_2}C_{k_3},
\label{43}
\end{equation}
where $C_k=(k-1)!! = (k-1) \times (k-3) \times (k-5) \times \cdots \times 1$ if $k=\text{even}$, being 0 if $k=\text{odd}$.

In the first stage of the calculations, the expansions  (\ref{40})--(\ref{39}) are inserted into  Eq.\ (\ref{24}). By equating the terms of degree $s$ in $g$ on both sides, we get 
\beqa
  \frac{\partial}{\partial x} M^{(s)}_{k_1+1,k_2,k_3} &+& \frac{\partial}{\partial y} M^{(s)}_{k_1,k_2+1,k_3} 
+
k_3 \sum_{i=0}^{\lbrack (s-1)/2 \rbrack} \left[ \frac{\partial u^{(2i+1)}}{\partial x} M^{(s-1-2i)}_{k_1+1,k_2,k_3-1}
\right. \nonumber \\&
+
&\left.
\frac{\partial u^{(2i+1)}}{\partial y} M^{(s-1-2i)}_{k_1,k_2+1,k_3-1} \right]- k_3 M^{(s-1)}_{k_1,k_2,k_3-1} = J^{(s)}_{\bf k},
\label{44}
\eeqa
where $ \lbrack \alpha \rbrack $ means the integer part of $\alpha$. According to Eq.\  (\ref{b5}), the right side of (\ref{44}) is 
\begin{equation}
\label{45}
J^{(s)}_{\bf k}={\sum_{{\bf k'},{\bf k''}}}^\dagger C_{{\bf k'},{\bf k''}}^{\bf k}\sum_{i=0}^s
M^{(s-i)}_{\bf k'}M^{(i)}_{\bf k''}.
\end{equation}

Equation (\ref{45}) is still very difficult to solve in general. First, it is  nonlinear and couples moments of degree $k$ and lower to moments of degree $k+1$. Second, the  coefficients  $M^{(s)}_{\bf k}$ are unknown functions of the spatial variables $x$ and $y$. To overcome the latter obstacle, we assume that a self-consistent solution of Eq.\ (\ref{44}) exists in which the  coefficients in the series defined by (\ref{40})--(\ref{39}) are \textit{polynomials} in $x$ and $y$. This ansatz is justified by the  results  obtained in the case of the planar geometry from the BGK model \cite{NOTE:3} and the Boltzmann equation \cite{NOTE:6}, as well as  in the case of the cylindrical geometry from the BGK model \cite{NOTE:10}.
More specifically, we assume that the coefficients of the hydrodynamic profiles are of the form
\beq
p^{(s)}=\sum_{i=1}^{s-1} p^{(s)}_{2i} r^{2i},
\label{46}
\eeq
\beq
T^{(s)}=\sum_{i=1}^{s} T^{(s)}_{2i} r^{2i},
\label{47}
\eeq
\beq
u^{(s)}=\sum_{i=1}^{s} u^{(s)}_{2i} r^{2i}.
\label{48}
\eeq
The task becomes more difficult when guessing the spatial dependence of the coefficients $M_{\bf k}^{(s)}$. It can be checked that the only representation that leads to consistent solutions of the hierarchy  (\ref{44}) is of the form
\begin{equation}
M^{(s)}_{\bf k} = \sum_{i=0}^{\sigma(s)} \sum_{j=0}^i \chi^{(s,i-j,j)}_{\bf k} x^{i-j}y^j,
\label{49}
\end{equation}
where  $\sigma(s=1)=1$ and  $\sigma(s)=2s$  for $s >1$. From Eqs.\ (\ref{n3}) it follows that $\chi^{(s,i,j)}_{k_1,k_2,k_3}=0$ if $s+k_3=\text{odd}$ or $i+k_1=\text{odd}$  or $j+k_2=\text{odd}$. Insertion of Eq.\ (\ref{49}) into Eq.\ (\ref{45}) shows that $J^{(s)}_{\bf k}$ has the same structure as $M^{(s)}_{\bf k}$, namely
\begin{equation}
J^{(s)}_{\bf k} = \sum_{i=0}^{\sigma(s)} \sum_{j=0}^i J^{(s,i-j,j)}_{\bf k} x^{i-j}y^j.
\label{49bis}
\end{equation}

The numerical coefficients $p^{(s)}_{2i}$, $T^{(s)}_{2i}$, $u^{(s)}_{2i}$ and $\chi^{(s,i,j)}_{\bf k}$ are  determined by consistency.
To do so, we insert Eqs.\ (\ref{46})--(\ref{49}) into Eq.\ (\ref{44}) and equate the coefficients of the terms $x^i y^j$ in both sides. This yields a hierarchy of \textit{linear} equations for the unknowns.
In the hierarchy, the coefficients of the moments of degree  $k+1$, $k$ and $k-1$ on the left-hand side are related to those of degree $k$, $k-1$, $k-2$, \ldots  on the right-hand side. In addition,  the coefficients  of the type $\chi^{(s,i+1,j)}_{k_1+1,k_2,k_3}$ and $\chi^{(s,i,j+1)}_{k_1,k_2+1,k_3}$ on the left-hand side  determine those of the type $\chi^{(s,i,j)}_{k_1,k_2,k_3}$ on the right-hand side. In spite of the intricacy of the resulting hierarchy, it can be recursively solved by following steps similar to those made in the case of the planar geometry \cite{NOTE:6}. 
To get explicit results to order $s=3$ it is necessary  to consider  the moment equations up to degree $k=8$.

The method is outlined in Appendix \ref{appB} and the main results are presented in the next Section.

\section{Discussion\label{sec5}}
\subsection{Hydrodynamic profiles}
By proceeding along the recursive scheme described in Appendix \ref{appB}, we have obtained the explicit spatial dependence of the coefficients in the series expansions (\ref{40})--(\ref{42}) through order $g^4$ and those of the series expansion (\ref{39}) for moments of second ($k=2$) and third ($k=3$) degrees through order $g^3$. Here we display the results to third order. Also, to ease the physical interpretation of the results, they are expressed in cylindrical coordinates  and in real units. The hydrodynamic profiles are
\beq
u_z(r)=u_0-\frac{\rho_0 r^2g}{4\eta_0}\left[1+\frac{1}{144} \frac{\rho_0^2 g^2}{\kappa_0 \eta_0 T_0}r^4+\zeta_u \left( \frac{mg}{k_B T_0}\right)^2r^2+\zeta_u' \frac{\rho_0 \eta_0^2 g^2}{p_0^3}\right]+\mathcal{O}(g^5),
\label{91}
\eeq
\beq
p(r)=p_0\left[1+\zeta_p\left( \frac{mg}{k_B T_0}\right)^2 r^2\right]+\mathcal{O}(g^4),
\label{92}
\eeq
\beq
T(r)=T_0 \left[1-\frac{1}{64} \frac{\rho_0^2 g^2}{\kappa_0 \eta_0 T_0}r^4+\zeta_T \left( \frac{mg}{k_B T_0}\right)^2r^2 \right]+\mathcal{O}(g^4),
\label{93}
\eeq
where $\zeta_u=\frac{4429}{12600}\simeq 0.35$, $\zeta_u'=24.322\ldots$, $\zeta_p=\frac{3}{10}$ and $\zeta_T=\frac{34}{175}\simeq 0.19$. 
We recall that the  subscript $0$ in a quantity represents its value  on the $z$-axis, i.e. at $r=0$.
Elimination of $r$ among Eqs.\ (\ref{91})--(\ref{93}) allows one to get the following nonequilibrium equation of state:
\beq
T=T_0-\frac{\eta_0}{4\kappa_0}(u_z-u_0)^2+
\frac{\zeta_T}{\zeta_p}\frac{T_0}{p_0}(p-p_0)+\mathcal{O}(g^4).
\label{5.4}
\eeq

The non-zero elements of the pressure tensor are
\beq
P_{rr}(r)=p_0\left[1 +\frac{1}{6}\zeta_p \left( \frac{mg}{k_B T_0}\right)^2 r^2-\zeta_{P}' \frac{\rho_0 \eta_0^2 g^2}{p_0^3}\right]+\mathcal{O}(g^4),
\label{94}
\eeq
\beq
P_{\phi\phi}(r)=p_0\left[1 +\frac{1}{2}\zeta_p \left( \frac{mg}{k_B T_0}\right)^2 r^2-\zeta_{P}' \frac{\rho_0 \eta_0^2 g^2}{p_0^3}\right]+\mathcal{O}(g^4),
\label{95}
\eeq
\beq
P_{zz}(r)=p_0\left[1 +\frac{7}{3}\zeta_p \left( \frac{mg}{k_B T_0}\right)^2 r^2+2\zeta_{P}' \frac{\rho_0 \eta_0^2 g^2}{p_0^3}\right]+\mathcal{O}(g^4),
\label{96}
\eeq
\beq
P_{rz}(r)=\frac{\rho_0 g}{2}r \left[1+\frac{1}{192}\frac{\rho_0^2 g^2}{ \kappa_0 \eta_0 T_0}r^4+\frac{\zeta_p-\zeta_T}{2} \left( \frac{mg}{k_B T_0}\right)^2 r^2 \right]+\mathcal{O}(g^5),
\label{97}
\eeq
where $\zeta_P'=1.7388\ldots$.
Finally, the components of the heat flux are
\beq
q_r(r)=\frac{1}{16} \frac{\rho_0^2 g^2}{\eta_0} r^3+\mathcal{O}(g^4),
\label{98}
\eeq
\beq
q_z(r)=-\frac{2}{5} \frac{\kappa_0 m g}{k_B} \left[\zeta_0 -\zeta_q\frac{\rho_0^2 g^2}{\kappa_0 \eta_0 T_0}r^4 -\zeta_q'\left( \frac{mg}{k_B T_0}\right)^2 r^2- \zeta_q'' \frac{\rho_0 \eta_0^2 g^2}{p_0^3} \right]+\mathcal{O}(g^5) ,
\label{99}
\eeq
where $\zeta_0=1$, $\zeta_q=\frac{37}{28}\simeq 0.29$, $\zeta_q'=3.2527\ldots$ and $\zeta_q''=86.415\ldots$.

Equations (\ref{91})--(\ref{99}) are the main results of this paper. They give the \textit{exact} profiles in the bulk region of the hydrodynamic  fields and their fluxes for a dilute gas of Maxwell molecules through third order in the external force.
Comparison of  Eqs.\ (\ref{2.20a}), (\ref{2.21b}) and (\ref{2.24.1})--(\ref{2.27}) with Eqs.\ (\ref{91})--(\ref{99}) show that the NS constitutive equations predict $\zeta_u=\zeta_u'=\zeta_p=\zeta_T=\zeta_P'=\zeta_0=\zeta_q=\zeta_q'=\zeta_q''=0$. These vanishing values of the coefficients summarize the main limitations of the NS description of the Poiseuille flow induced by an external force. 
Equation (\ref{99}) shows that $q_z\neq 0$, even to
first order in $g$ ($\zeta_0\neq 0$), despite the absence of a thermal gradient along the longitudinal direction. This represents an obvious breakdown of  Fourier's law and is in fact a Burnett-order effect \cite{NOTE:6}.
To second order in $g$ the hydrostatic pressure is not uniform but increases radially ($\zeta_p>0$); analogously, there are normal stress differences, so that $P_{rr}<P_{\phi\phi}<p<P_{zz}$. These are non-Newtonian effects. 

Also to second order in $g$, there exists a positive quadratic term ($\zeta_T>0$) in addition to the negative quartic term in the temperature profile. The former term is responsible for the fact that the temperature  has a local minimum rather than a maximum at $r=0$. The maximum temperature is located at a distance $r_{\text{max}}=\sqrt{32\zeta_T}\ell_0\simeq 2.49\ell_0$ (independent of $g$), where $\ell_0\equiv ( \eta_0 \kappa_0 T_0 )^{1/2}/p_0=\sqrt{\frac{2}{5}\text{Pr}}(v_0/\nu_0)$ is the (local) mean free path at $r=0$; the relative difference between the maximum temperature and the temperature on the cylinder axis is $(T_{\text{max}}-T_0)/T_0=16 \zeta_T^2 ( \ell_0/h_0)^2\simeq 0.604( \ell_0/h_0)^2$, where we recall that $h_0\equiv k_BT_0/mg$. This result represents a dramatic violation of Fourier's law: while  the heat flows radially outwards, the temperature \text{increases} from $r=0$ to $r=r_{\text{max}}$. Therefore, within the region $0\leq r\leq r_{\text{max}}$, the heat flows from the colder to the hotter layers. This paradoxical effect, which is beyond the Burnett description \cite{NOTE:5}, does not violate the conservation of energy (\ref{2.16}) since the radial increase of $rq_r$ (which tends to cool the gas) is exactly compensated for by the viscous heating term $rP_{rz}\partial u_z/\partial r$. It is interesting to note that from Eqs.\ (\ref{93}) and (\ref{98}) one can obtain
\beq
-\kappa \frac{\partial T}{\partial r}=q_r-\frac{1}{9}r_{\text{max}}^2\nabla^2 q_r+\mathcal{O}(g^4),
\label{x1}
\eeq
where $\nabla^2 X=r^{-1}\partial(r\partial X/\partial r)/\partial r$ is the Laplacian in cylindrical coordinates. Equation (\ref{x1}) is an extension of Fourier's law (\ref{2.21a}) showing that the sign of the thermal gradient results from a competition between the radial component of the heat flux and its Laplacian \cite{NOTE:9}. The latter dominates for $r<r_{\text{max}}$ and so $\partial T/\partial r>0$ in that region.

Third-order contributions appear in $u_z$ and $P_{rz}$ but they do not have a qualitatively important influence. They are responsible for a spatial variation of the flow velocity $u_z$ and the shear stress $P_{rz}$  more pronounced than expected from the NS equations. The same happens for the 
mass rate of flow defined by Eq.\ (\ref{new4}). Insertion of the hydrodynamic profiles (\ref{91})--(\ref{93}) yields
\beqa
\dot{M}(a)&=&\frac{\pi \rho_0^2 g a^4}{8\eta_0}\left[1+\frac{5}{384}\frac{\rho_0^2g^2}{\eta_0\kappa_0T_0}
a^4+\frac{4\zeta_u+\zeta_p-\zeta_T}{3} \left( \frac{mg}{k_B T_0}\right)^2a^2+\zeta_u' \frac{\rho_0 \eta_0^2 g^2}{p_0^3}\right]\nonumber\\
&&+\mathcal{O}(g^5).
\label{new7}
\eeqa
Comparison with Eq.\ (\ref{new5}) shows that the mass rate of flow grows with the radius $a$ more rapidly than predicted by the NS approximation, but otherwise the qualitative behavior is similar in both descriptions.

\subsection{Generalized constitutive equations}
Let us use Eqs.\ (\ref{91})--(\ref{99}) to rewrite the fluxes under the form of \textit{generalized} constitutive equations. We begin with the shear stress:
\beqa
P_{rz}&=&-\eta\frac{\partial u_z}{\partial r}\biggl[1-\frac{\zeta_u'}{15\zeta_T}\frac{\ell ^2}{T}\nabla^2 T-\frac{4}{15}\left(\frac{\zeta_u'}{15\zeta_T}+8\zeta_u+6\zeta_T-2\zeta_p\right)
\nonumber\\
&&\times
\frac{\ell ^2}{k_BT/m}\left(\frac{\partial u_z}{\partial r}\right)^2\biggr]+\mathcal{O}(g^5).
\label{y1}
\eeqa
Here $\eta\propto T$ is the local shear viscosity at $r$, so $\eta-\eta_0\propto T-T_0=\mathcal{O}(g^2)$. Analogously, $\ell=\left(\eta\kappa T\right)^{1/2}/p$ is the local mean free path and $\ell-\ell_0=\mathcal{O}(g^2)$. In Eq.\ (\ref{y1}) the NS term is of order $g$, while  the  super-Burnett terms (i.e. of third order in the hydrodynamic gradients) are of order $g^3$. 
 The normal stresses can be written as
\beq
P_{rr}=p\left[1+\frac{10\zeta_p}{3}\frac{\ell ^2}{T}\nabla^2T-\left(\frac{\zeta_P'}{15\zeta_p}+\frac{10\zeta_T}{3}\right)\frac{\ell ^2}{p}\nabla^2 p\right]+\mathcal{O}(g^4),
\label{y2}
\eeq
\beq
P_{\phi\phi}=p\left[1+2\zeta_p\frac{\ell ^2}{T}\nabla^2T-\left(\frac{\zeta_P'}{15\zeta_p}+2 \zeta_T\right)\frac{\ell ^2}{p}\nabla^2 p\right]+\mathcal{O}(g^4),
\label{y3}
\eeq
\beq
P_{zz}=p\left[1-\frac{16\zeta_p}{3}\frac{\ell ^2}{T}\nabla^2T+\left(\frac{2\zeta_P'}{15\zeta_p}+\frac{16\zeta_T}{3}\right)\frac{\ell ^2}{p}\nabla^2 p\right]+\mathcal{O}(g^4).
\label{y4}
\eeq
Thus, in order to obtain the normal stresses to order $g^2$ we need to include at least the Burnett contributions in the Chapman--Enskog expansion.

 The radial component of the heat flux can be expressed as
\beq
q_r=-\kappa \frac{\partial}{\partial r}\left(T+4\zeta_T\ell^2 \nabla^2 T\right)+\mathcal{O} (g^4).
\label{y5}
\eeq
In this case, both the NS term and the  super-Burnett term are of order $g^2$. As for the longitudinal heat flux, one has
\beqa
q_z &=&\frac{\eta^2}{2\rho}\left(\nabla^2 u_z\right)\left(\theta_4-\frac{\zeta_u'+\zeta_q''}{5\zeta_T}\nabla^2 T\right)
+\frac{2}{5}nk_B\ell ^2 \frac{\partial u_z}{\partial r}\nonumber\\
&&\times\left(\theta_5\frac{\partial T}{\partial r}  +\frac{\theta_3}{3}\frac{T}{p}\frac{\partial p}{\partial r}
+\mu \ell ^2\frac{\partial}{\partial r}\nabla^2T\right)+\mathcal{O} (g^5),
\label{y6}
\eeqa
where $\theta_3=-3$, $\theta_4=3\zeta_0=3$, $\theta_5=\frac{1}{2}({1+64\zeta_q})=\frac{39}{4}$ and
\beq
\mu=\frac{\zeta_u'+\zeta_q''}{15\zeta_T}+16\zeta_u+2\zeta_T(7+64\zeta_q)+4\zeta_q'-8\zeta_p\simeq 64.141\ldots .
\label{y7}
\eeq
The Burnett term headed by $\theta_4$ is of of first order in $g$, while the Burnett terms with $\theta_3$ and $\theta_5$ are of third order. In addition, there are super-super-Burnett terms that are also of order $g^3$.

Equations (\ref{y1})--(\ref{y6}) can be written in other equivalent forms by taking into account that the hydrodynamic gradients are not independent in our problem. For instance, 
\beq	
\frac{m}{k_BT }\left(\frac{\partial u_z}{\partial r}\right)^2=\frac{15}{4}\left(\frac{\zeta_T}{\zeta_p}\frac{1}{p}\nabla^2p-\frac{1}{T}\nabla^2T\right)
+\mathcal{O} (g^4)
\label{y8},
\eeq
\beq
\nabla^2 p=-4\zeta_p \frac{p}{T}\ell^2 \nabla^4 T+\mathcal{O}(g^4)=\frac{16}{15}\zeta_p\frac{mp}{k_BT}\ell^2\left(\nabla^2 u_z\right)^2+\mathcal{O}(g^4).
\label{y9}
\eeq
In particular, Eq.\ (\ref{y9}) implies that the Burnett term $\nabla^2 p$ and the super-super-Burnett terms $\nabla^4 T$ and $\left(\nabla^2 u_z\right)^2$ are of order $g^2$. Therefore, in order to obtain correctly the profiles (\ref{91})--(\ref{99}) from the Chapman--Enskog method one would actually need to go beyond the apparent orders in the gradients of Eqs.\ (\ref{y1})--(\ref{y6}).

The previous analysis shows that the Chapman--Enskog expansion and the expansion in powers of $g$ in the Poiseuille flow are quite different, even though  the gradients are induced by the external force. Both expansions require a weak field, namely $h_0\gg \ell_0$. On the other hand, while the Chapman--Enskog expansion requires that $\ell_0^2T_0^{-1}\nabla^2 T\ll 1$, i.e. $r\gg \ell_0$, the expansion in powers of $g$ remains valid for distances smaller than or of the order of the mean free path. Since the non-monotonic behavior of the temperature profile occurs for $0\leq r\leq r_{\text{max}}\sim \ell_0$, this effect is neglected by the NS description.

\subsection{Comparison with the BGK solution}
\begin{table}
\caption{Numerical values of the coefficients appearing in Eqs.\ (\protect\ref{91})--(\protect\ref{99}), as obtained from the Boltzmann equation for Maxwell molecules (BM) and for from the BGK model for Maxwell molecules. The ratio between the BM and BGK values are displayed in the third column.\label{tab1}}
\begin{center}
\begin{tabular}{cccc}
\hline 
$\zeta$      & BM   & BGK                         & BM/BGK       \\ \hline
$\zeta_u$ & $\frac{4429}{12600}$ 	&$\frac{89}{200}$	&$0.79$  		\\ 
$\zeta_u'$ & 	$24.322$	& $64$		&$0.38$  		\\ 
$\zeta_p$ & $\frac{3}{10}$	& $\frac{3}{10}$		&$1$  		\\ 
$\zeta_T$ & $\frac{34}{175}$	& $\frac{7}{50}$	&$1.38$  	   \\ 
$\zeta_{P}'$ &$1.7388$ 	& $\frac{92}{25}$	&$0.47$  		\\ 
$\zeta_0$ &$1$ 	&$1$ 	&$1$  		\\ 
$\zeta_{q}$ & $\frac{37}{128}$	&$\frac{15}{64}$ 	&$1.23$  \\	
$\zeta_{q}'$ &$3.2527$ 	& $\frac{209}{50}$	&$0.78$  		\\ 
$\zeta_{q}^{''}$ & $86.415$	& $\frac{5432}{25}$	&$0.39$ \\ 		
\hline 
\end{tabular}
\end{center}
\end{table}
Once we have discussed the limits of validity of the NS theory (which otherwise provides a powerful tool to study the macroscopic state of many nonequilibrium flows) in our Poiseuille problem, it is worthwhile wondering whether the BGK model kinetic equation succeeds in capturing the main qualitative features of the Boltzmann solution.
The solution of the BGK model for the cylindrical Poiseuille flow was obtained by Tij and Santos \cite{NOTE:10} to fourth order in $g$ for a general class of power-law repulsive potentials. The results agree exactly with the structure of Eqs.\ (\ref{91})--(\ref{99}). On the other hand, most of the  coefficients $\zeta$'s have different numerical values. The failure of the BGK model to reproduce the exact Boltzmann values for those coefficients is not surprising, given the simplicity of the model, which assumes that the practical effect of collisions is to make the distribution function relax to local equilibrium with a single characteristic rate $\nu$. In fact, it is well known \cite{NOTE:12} that the BGK model yields a value ($\text{Pr}=1$) for the Prandtl number $\text{Pr}\equiv 5k_B\eta/2m\kappa$ different from the correct one ($\text{Pr}=\frac{2}{3}$). 
Table \ref{tab1} compares the numerical values of the coefficients $\zeta$'s given by the Boltzmann equation for Maxwell molecules (BM) with those obtained from the BGK model.
Except for $\zeta_p$ and $\zeta_0$, the coefficients are different in both theories, the ratio being bounded between $0.38$ and $1.38$.
For instance, the location $r_\text{max}$ of the maximum temperature is in the BGK model 15\% smaller than in the Boltzmann equation and the temperature difference $T_{\text{max}}-T_0$ in the BGK model is almost half the Boltzmann value.
A comparison of the coefficients appearing in the hydrodynamic fields to fourth order in $g$ [not shown in Eqs.\ (\ref{92}) and (\ref{93}), but given in Appendix \ref{appB}] shows that the discrepancies between the BGK and Boltzmann equations increase with the  power of $g$. In particular,  the series obtained from the BGK model seem to diverge more quickly than those obtained from the Boltzmann equation. 

\begin{figure}[tbp]
 \includegraphics[width=.85 \columnwidth]{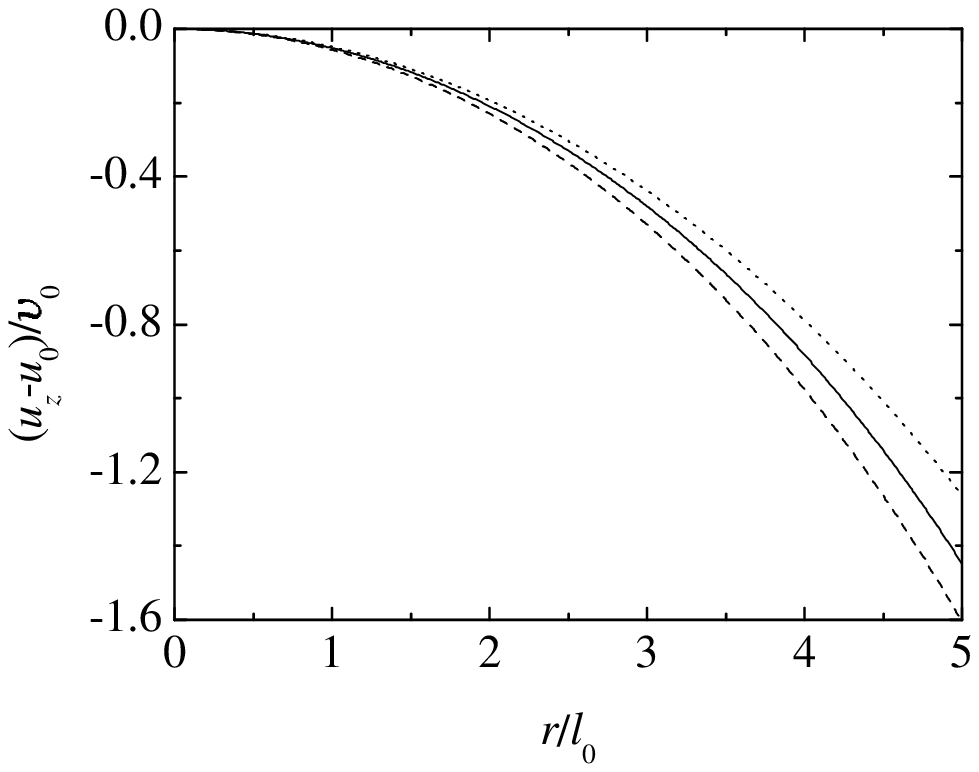}
 \caption{Profile of the flow velocity $u_z(r)$ for $g=0.1v_0^2/\ell_0$, as obtained from the Navier--Stokes (NS) equations (dotted line), the Boltzmann equation for Maxwell molecules (BM) (solid line) and the BGK model (dashed line).\label{fig1}}
 \end{figure}
\begin{figure}[tbp]
 \includegraphics[width=.85 \columnwidth]{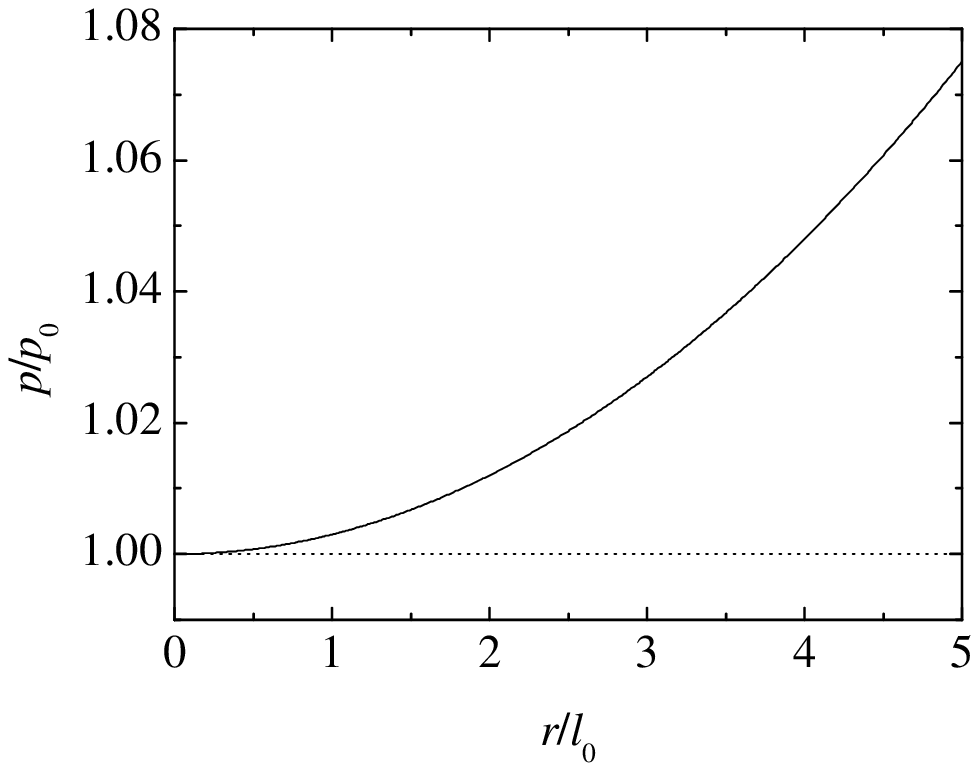}
 \caption{Same as in Fig.\ \protect\ref{fig1}, but for the hydrostatic pressure  $p(r)$. Note that the  BM and BGK curves coincide.\label{fig2}}
 \end{figure}
\begin{figure}[tbp]
 \includegraphics[width=.85 \columnwidth]{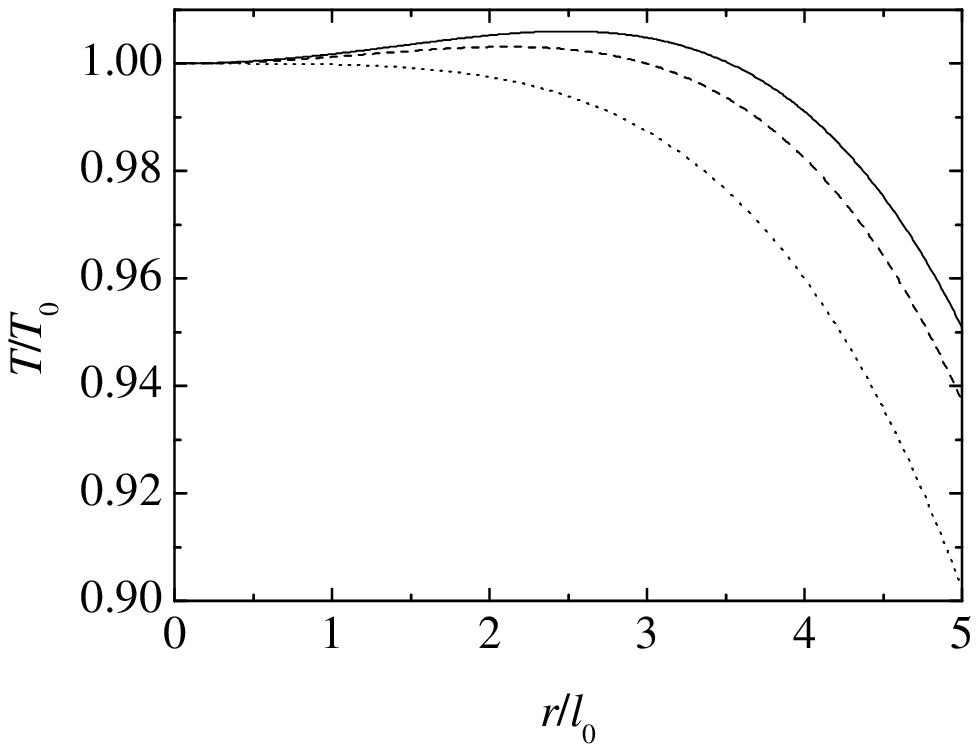}
 \caption{Same as in Fig.\ \protect\ref{fig1}, but for the temperature $T(r)$.\label{fig3}}
 \end{figure}
\begin{figure}[tbp]
 \includegraphics[width=.85 \columnwidth]{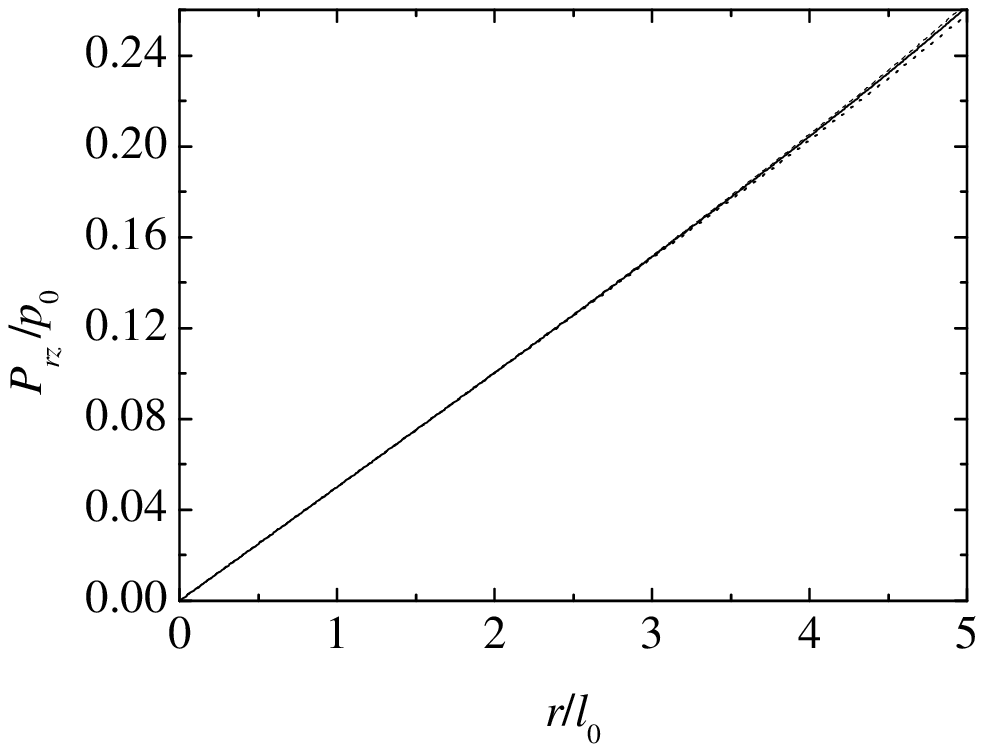}
 \caption{Same as in Fig.\ \protect\ref{fig1}, but for the shear stress  $P_{rz}(r)$. The NS and BGK curves are slightly below and above, respectively, the BM curve.\label{fig4}}
 \end{figure}
\begin{figure}[tbp]
 \includegraphics[width=.85 \columnwidth]{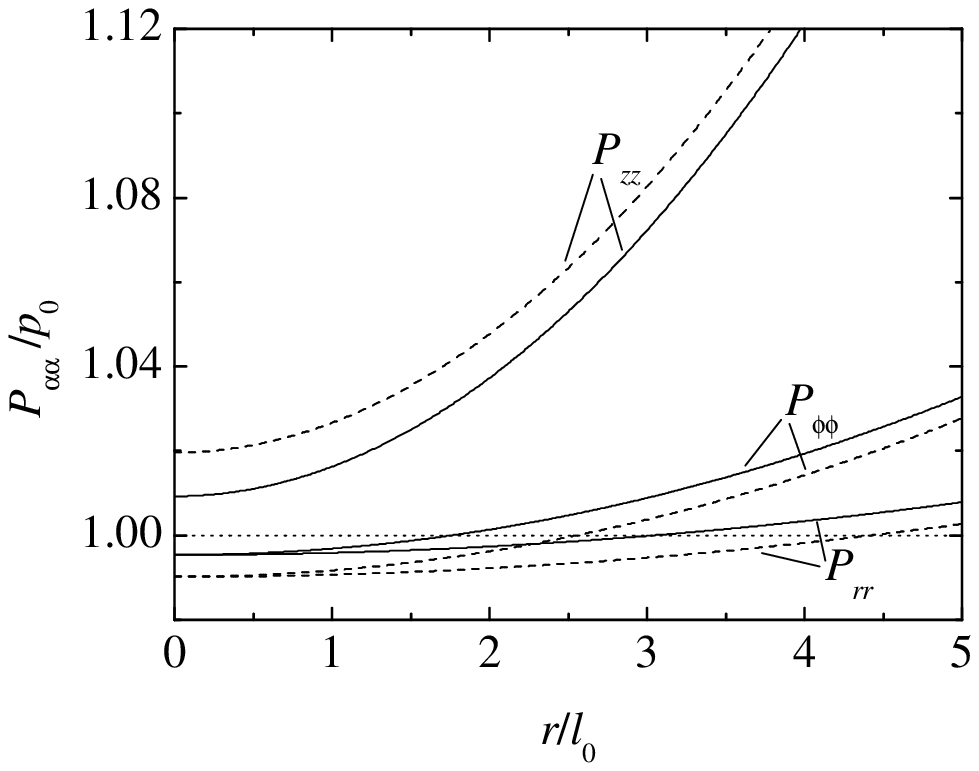}
 \caption{Same as in Fig.\ \protect\ref{fig1}, but for the normal stresses  $P_{rr}(r)$, $P_{\phi\phi}(r)$ and $P_{zz}(r)$. Note that in the NS approximation $P_{rr}=P_{\phi\phi}=P_{zz}=p$.\label{fig5}}
 \end{figure}
\begin{figure}[tbp]
\includegraphics[width=.85 \columnwidth]{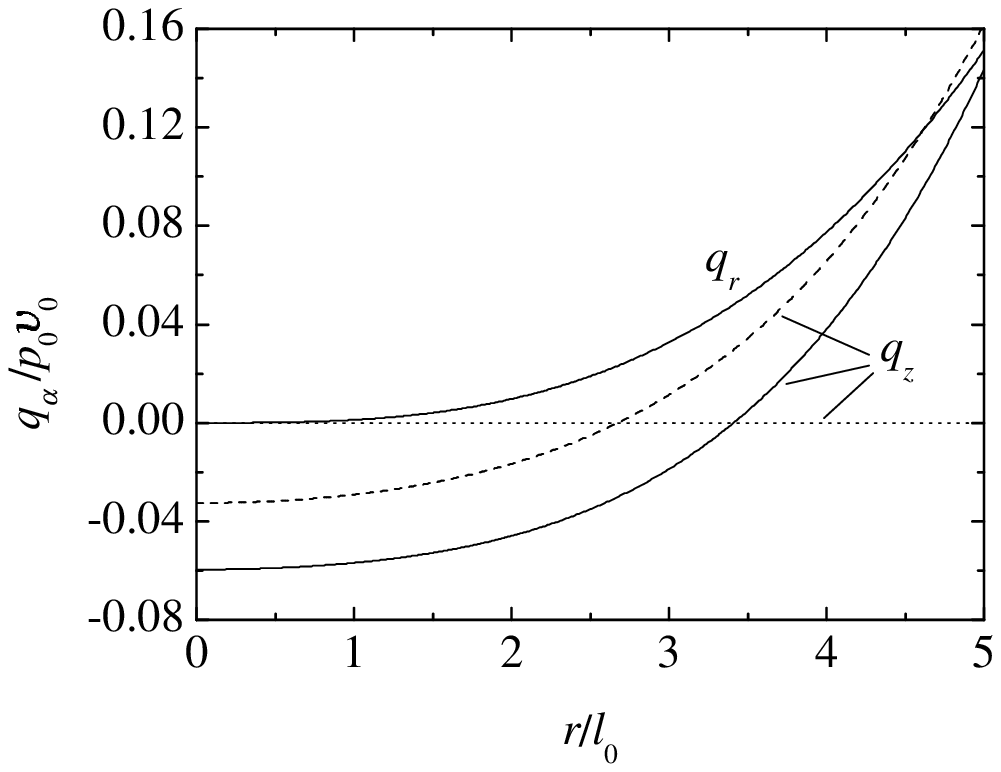}
 \caption{Same as in Fig.\ \protect\ref{fig1}, but for the components of the heat flux. Note that the NS, BM and BGK curves coincide for the radial component $q_{r}(r)$.\label{fig6}}
 \end{figure}
In order to illustrate graphically the differences between the NS, BGK and BM descriptions, we plot  in Figs.\ \ref{fig1}--\ref{fig6} the
hydrodynamic and flux profiles for the case $g=0.1v_0^2/\ell_0$
[which corresponds to $h_0/\ell_0=10$],  when only terms through third order in $g$ are retained.
Of course, higher order terms are not  expected to be negligible for that
particular high value of the external field \cite{note}, especially as one departs from the axis, as discussed in  Subsection \ref{subsec5.5}. However,  the retained terms are sufficient to show the qualitative differences in the predictions of the three approaches, so that the main aim of Figs.\ \ref{fig1}--\ref{fig6} is to highlight those differences. 
It must be born in mind that the BGK curves in Figs.\ \ref{fig1}--\ref{fig6} correspond to Eqs.\ (\ref{91})--(\ref{99}) with the BGK values for the coefficients $\zeta$'s (cf. Table \ref{tab1}) but with the correct value $\text{Pr}=\frac{2}{3}$ of the Prandtl number.
As expected, the BGK model is in qualitative agreement with the BM results. 

\subsection{Comparison with the planar Poiseuille flow}
Let us now add a few comments on the similarities between the cylindrical and the planar Poiseuille flows. The solution of the Boltzmann equation for Maxwell molecules to order $g^2$ in the planar geometry \cite{NOTE:6} yields profiles analogous to those of Eqs.\ (\ref{91})--(\ref{99}). In particular, the temperature profile is 
\beq
T(y)=T_0 \left[1-\frac{1}{12} \frac{\rho_0^2 g^2}{\kappa_0 \eta_0 T_0}y^4+\zeta_T \left( \frac{mg}{k_B T_0}\right)^2y^2 \right]+\mathcal{O}(g^4),
\label{93bis}
\eeq
where now $\zeta_T=1.0153$. As a consequence, the temperature has a maximum with $(T_{\text{max}}-T_0)/T_0=3\zeta_T^2(\ell_0/h_0)^2\simeq 3.09 (\ell_0/h_0)^2$ at $|y|=y_{\text{max}}=\sqrt{6\zeta_T}\ell_0\simeq 2.47\ell_0$. Therefore, the maximum temperature occurs at a separation $y_{\text{max}}$ from the  middle plane practically equal  to the radial distance $r_{\text{max}}$ of the cylindrical case, but the relative temperature change $(T_{\text{max}}-T_0)/T_0$ is about five times larger in the former case than in the latter. Interestingly, this effect is accurately captured by the BGK model \cite{NOTE:10}, as explained by the fact that in the planar Poiseuille flow one has \cite{NOTE:3,NOTE:6} $\zeta_p^{\text{BM}}/\zeta_p^{\text{BGK}}=1$, $\zeta_T^{\text{BM}}/\zeta_T^{\text{BGK}}=1.0153/(19/25)\simeq 1.34$, ${\zeta_P'}^{\text{BM}}/{\zeta_P'}^{\text{BGK}}=6.2602/(306/25)\simeq 0.51$, $\zeta_0^{\text{BM}}/\zeta_0^{\text{BBGK}}=1$, in close agreement with the corresponding ratios in the cylindrical Poiseuille flow (see Table \ref{tab1}).

\subsection{Boundary conditions}
\label{subsec5.5}
When solving the Boltzmann hierarchy of moments [cf.\ Eq.\ (\ref{24})] by a perturbation expansion in powers of gravity we have taken the hydrodynamic quantities at the axis ($r=0$) as reference values. In so doing, we have avoided the need of imposing specific boundary conditions. The price to be paid is that it is difficult to determine how far from the axis the profiles (\ref{91})--(\ref{99}) remain valid. Consider for instance the temperature profile (\ref{93}). The first neglected term $\mathcal{ O}(g^4)$ is a polynomial in $r$ of degree 8 [cf.\ Eq.\ (\ref{67})]. This term can actually be neglected versus the retained terms only if ${g^*}^4(r/\ell_0)^8\ll {g^*}^2(r/\ell_0)^4$, i.e. if 
\beq
{g^*}(r/\ell_0)^2\ll 1. 
\label{new10}
\eeq
The same conclusion is obtained from the velocity profile (\ref{91}) and the pressure profile (\ref{92}). If $r\sim\ell_0$, condition (\ref{new10}) is equivalent to $g^*\ll 1$. On the other hand, a much stronger condition $g^*\ll \text{Kn}^2$ is needed to extend the profiles (\ref{91})--(\ref{99}) to distances $r$ comparable with the pipe radius $R$. 
In order to find the hydrodynamic fields for $r\sim R$ when condition (\ref{new10}) does not hold, one would need to solve the full Boltzmann equation with the appropriate (e.g. diffuse reflection) boundary conditions at $r=R$ corresponding to a given wall temperature $T_w$. 
This has been done by Aoki et al.\ \cite{ATN02} in the case of the BGK equation for the planar geometry. They derived the hydrodynamic equations and the matching conditions corresponding to the normal (or Hilbert) solution by an asymptotic expansion in powers of the Knudsen number $\text{Kn}$ and for values  of the Froude number of order $\text{Fr}\sim 1/\text{Kn}$.
In view of Eq.\ (\ref{new9}), this implies that $g^*\sim \text{Kn}^2$.
The successive hydrodynamic equations had to be solved numerically, but Aoki et al.\ verified that a Taylor series expansion around the center of the gap agreed with the perturbation solution found in Ref.\ \cite{NOTE:3}. 

An asymptotic method  analogous to that worked out by Aoki et al.\ \cite{ATN02} would be much more difficult to carry out in the case of the Boltzmann equation for the cylindrical geometry. Nevertheless, it is natural to expect that the results derived in this paper would be recovered by performing a Taylor series expansion around $r=0$ of the solution for 
$g^*\sim \text{Kn}^2$. This in addition would provide the matching conditions for the profiles (\ref{91})--(\ref{93}) in the region $r/\ell_0\sim {g^*}^{-1/2}\sim \text{Kn}^{-1}$.

\section{Conclusion\label{sec6}}

In this work, we have studied the (laminar) stationary Poiseuille flow in a cylindrical pipe  induced by an external force  $m\mathbf{g}$. 
We have considered a dilute gas of Maxwell molecules and have analyzed the hierarchy of moment equations associated with the nonlinear Boltzmann equation.
A consistent solution has been found under the form of a perturbation expansion in powers of $g$ through third order. In principle, the method can be pushed to higher orders but, not only the algebraic intricacy of the solution scheme grows rapidly, but the results would not be of practical use given the suspected asymptotic character of the series.
The discussion of the results has focused on the most relevant physical quantities, namely the hydrodynamic profiles (flow velocity, pressure and temperature) and the associated fluxes (pressure tensor and heat flux).
 The results obtained here show once again that the Navier--Stokes (NS) theory remains unable to envisage the correct hydrodynamic profiles, even at a qualitative level, for distances smaller than or of the order of the mean free path.
The most important limitation of the NS
description is that it predicts a monotonically decreasing temperature as
one moves apart from the cylinder axis. In contrast, our solution to the Boltzmann equation  shows that the temperature has a local minimum ($T=T_0$) at the
axis ($r=0$) and reaches a maximum value $T=T_{\text{max}}\simeq T_0\left[1+0.6(\ell_0/h_0)^2\right]$ at a distance
 from the center  $r=r_{\text{max}}\simeq 2.5\ell_0$ of the order of the mean free path $\ell_0$. 
In addition, a longitudinal component of the heat flux exists ($q_z\neq 0$) in the absence
of gradients along the longitudinal direction and normal stress differences ($P_{rr}<P_{\phi\phi}<p<P_{zz}$) are present.
 On the other hand, the BGK model provides results qualitatively similar to those found in this paper \cite{NOTE:10}. Notwithstanding this,  the profiles predicted by the BGK model require some corrections from a quantitative point of view, as expected. In particular, the BGK model underestimates the maximum value $T_{\text{max}}$ of the temperature, as well as its location $r_{\text{max}}$. From that point of view, one might say that the BGK model  deviates from the NS predictions less than the Boltzmann equation, at least for Maxwell molecules.
The same conclusion arises from a comparison between the Boltzmann and BGK results for the planar Poiseuille flow \cite{NOTE:3,NOTE:6}.
Nevertheless, it is worth noting that the BGK model is accurate in describing the effect of the geometry of the Poiseuille flow on the profiles.
For instance, it correctly predicts that in the planar case the maximum temperature is located at a separation $y_{\text{max}}$ from the  middle plane practically equal  to the radial distance $r_{\text{max}}$ of the cylindrical case, although the relative temperature change $(T_{\text{max}}-T_0)/T_0$ is about five times larger in the former case than in the latter.

Although we have not considered interactions different from Maxwell's, we expect the results found here to remain essentially valid, except that the numerical values of the coefficients may change. This expectation is supported by the results obtained with the BGK kinetic model \cite{NOTE:10}. In that case one finds that the coefficients $\zeta_p$, $\zeta_T$, $\zeta_P'$ and $\zeta_0$ are universal, while the coefficients $\zeta_u$, $\zeta_u'$, $\zeta_q$, $\zeta_q'$ and  $\zeta_q''$ change by 5\%, 0.5\%, 3\%, 0.7\% and 0.8\%, respectively, when going from Maxwell molecules to hard spheres.

Let us conclude by asserting that the stationary Poiseuille flow driven  by an external force, both in a slab and in a pipe,  is a good and conceptually simple example in fluid dynamics showing the limitations of a purely continuum theory in contrast to a kinetic theory approach.
Next, we want to emphasize the important role played by the BGK model kinetic equation to pave the way towards a more fundamental theory based on the Boltzmann equation.

\ack
A.S. gratefully acknowledges partial support from the Ministerio de
Ciencia y Tecnolog\'{\i}a
 (Spain) through grant No.\ BFM2001-0718.

\appendix
\section{Cylindrical coordinates}
\label{appA}
The relationships between the cylindrical and Cartesian
components of a vector ${\bf A}$ and a tensor ${\sf B}$ are
\beq
\left(
\begin{array}{c}
A_r\\
A_\phi\\
A_z
\end{array}
\right)=
{\sf U}\cdot
\left(
\begin{array}{c}
A_x\\
A_y\\
A_z
\end{array}
\right),
\label{2.17}
\eeq
\beq
\left(
\begin{array}{ccc}
B_{rr}&B_{r\phi}&B_{rz}\\
B_{\phi r}&B_{\phi\phi}&B_{\phi z}\\
B_{zr}&B_{z\phi}&B_{zz}
\end{array}
\right)=
{\sf U}\cdot
\left(
\begin{array}{ccc}
B_{xx}&B_{xy}&B_{xz}\\
B_{yx}&B_{yy}&B_{y z}\\
B_{zx}&B_{zy}&B_{zz}
\end{array}
\right)
\cdot {\sf U}^\dagger,
\label{2.18}
\eeq
where
\beq
{\sf U}=
\left(
\begin{array}{ccc}
x/r&y/r&0\\
-y/r&x/r&0\\
0&0&1
\end{array}
\right)
\label{2.19}
\eeq
is a unitary matrix and ${\sf U}^\dagger$ is its transpose.

In general, the divergence of a vector $\mathbf{A}$ in cylindrical coordinates is \cite{A89}
\beq
\nabla \cdot \mathbf{A}=\frac{1}{r}\frac{\partial}{\partial r}\left(rA_r\right)+\frac{1}{r}\frac{\partial}{\partial \phi}A_\phi+\frac{\partial}{\partial z}A_z.
\label{B1}
\eeq
The cylindrical components of the divergence of a tensor $\mathsf{B}$ are \cite{A89}
\beq
\left(\nabla\cdot\mathsf{B}\right)_r=\frac{1}{r}\frac{\partial}{\partial r}\left(rB_{rr}\right)+\frac{1}{r}\frac{\partial}{\partial \phi}B_{\phi r}+\frac{\partial}{\partial z}B_{zr}-\frac{1}{r}B_{\phi\phi},
\label{B2}
\eeq
\beq
\left(\nabla\cdot\mathsf{B}\right)_\phi=\frac{1}{r}\frac{\partial}{\partial r}\left(rB_{r\phi}\right)+\frac{1}{r}\frac{\partial}{\partial \phi}B_{\phi \phi}+\frac{\partial}{\partial z}B_{z\phi}+\frac{1}{r}B_{\phi r},
\label{B3}
\eeq
\beq
\left(\nabla\cdot\mathsf{B}\right)_z=\frac{1}{r}\frac{\partial}{\partial r}\left(rB_{r z}\right)+\frac{1}{r}\frac{\partial}{\partial \phi}B_{\phi z}+\frac{\partial}{\partial z}B_{zz}.
\label{B4}
\eeq

In the Poiseuille problem the cylindrical components of the flow velocity $\mathbf{u}$, the heat flux $\mathbf{q}$ and the pressure tensor $\mathsf{P}$ depend  on the radial  variable $r$ only. Consequently,
\beq
\nabla\cdot \mathsf{q}=\frac{1}{r}\frac{\partial}{\partial r}\left(rq_r\right),
\label{B5}
\eeq
\beq
\left(\nabla\cdot\mathsf{P}\right)_r=\frac{1}{r}\frac{\partial}{\partial r}\left(rP_{rr}\right)-\frac{1}{r}P_{\phi\phi},
\label{B6}
\eeq
\beq
\left(\nabla\cdot\mathsf{P}\right)_\phi=0,
\label{B7}
\eeq
\beq
\left(\nabla\cdot\mathsf{P}\right)_z=\frac{1}{r}\frac{\partial}{\partial r}\left(rP_{r z}\right),
\label{B8}
\eeq
\beq
\mathsf{P}:\nabla \mathbf{u}=P_{rz}\frac{\partial u_z}{\partial r}.
\label{B9}
\eeq
In Eq.\ (\ref{B7}) we have taken into account that, by symmetry, $P_{r\phi}=P_{\phi r}=0$.

\section{Solution of the hierarchy of moment equations}
\label{appB}
In this Appendix we outline the steps needed to solve the hierarchy (\ref{44}) with  the profiles (\ref{46})--(\ref{49}). To do so, we proceed sequentially from a given order $s$ to the next order $s+1$.  It is also necessary to use the consistency relation
\begin{equation}
p=\frac {1}{3} \left( M_{2,0,0}+M_{0,2,0}+M_{0,0,2} \right)
\label{64}
\end{equation}
to any order in $g$.
Let us start from the first-order equations.
\subsection{First order in $g$  $(s=1)$}

Setting $s = 1$,  Eq.\ (\ref{44}) becomes
\beqa
\frac{\partial }{\partial x} M^{(1)}_{k_1+1,k_2,k_3} &+& \frac{\partial}{\partial y} M^{(1)}_{k_1,k_2+1,k_3} + k_3 \frac{\partial u^{(1)}}{\partial x} M^{(0)}_{k_1+1,k_2,k_3-1} \nonumber\\
&+&
k_3 \frac{\partial u^{(1)}}{\partial y} M^{(0)}_{k_1,k_2+1,k_3-1}- k_3 M^{(0)}_{k_1,k_2,k_3-1} = J^{(1)}_{\bf k}.
\label{50}
\eeqa
Next, inserting (\ref{48}) and (\ref{49}) into (\ref{50}), we obtain
\beqa
\chi^{(1,1,0)}_{k_1+1,k_2,k_3}&+&\chi^{(1,0,1)}_{k_1,k_2+1,k_3}+2 k_3 M^{(0)}_{k_1+1,k_2,k_3-1} u^{(1)}_2 x
+2 k_3 M^{(0)}_{k_1,k_2+1,k_3-1} u^{(1)}_2 y\nonumber\\
&-&k_3 M^{(0)}_{k_1,k_2,k_3-1}=J^{(1,0,0)}_{\bf k}+J^{(1,1,0)}_{\bf k} x + J^{(1,0,1)}_{\bf k}y,\nonumber\\
\label{51}
\eeqa
where  $J^{(s,i,j)}_{\bf k}$ is defined by Eq.\ (\ref{49bis}).
Equating the coefficients of the same degree in $x$ and $y$, Eq.\ (\ref{51}) decouples into the following set of equations:
\beq
2 k_3 M^{(0)}_{k_1+1,k_2,k_3-1} u^{(1)}_2 = J^{(1,1,0)}_{k_1,k_2,k_3},
\label{52}
\eeq
\beq
2 k_3 M^{(0)}_{k_1,k_2+1,k_3-1} u^{(1)}_2 = J^{(1,0,1)}_{k_1,k_2,k_3}, 
\label{53}
\eeq
\beq
\chi^{(1,1,0)}_{k_1+1,k_2,k_3}+ \chi^{(1,0,1)}_{k_1,k_2+1,k_3} - k_3 M^{(0)}_{k_1,k_2,k_3-1}=J^{(1,0,0)}_{k_1,k_2,k_3}.
\label{54}
\eeq

Let us start with the moments of small degree ($k=2$). For example, by taking $\lbrace k_1,k_2, k_3 \rbrace = \lbrace 1,0,1 \rbrace $ in (\ref{52}) and $\lbrace k_1,k_2, k_3 \rbrace = \lbrace 0,1,1 \rbrace $ in (\ref{53},) we obtain
\beq
2 u^{(1)}_2=-\chi^{(1,1,0)}_{1,0,1}   ,
\label{55}
\eeq
\beq
2 u^{(1)}_2=-\chi^{(1,0,1)}_{0,1,1} .
\label{56}
\eeq
Now we insert the above equations into Eq.\  (\ref{54}) with $\lbrace k_1,k_2, k_3 \rbrace = \lbrace 0,0,1 \rbrace $. This allows us to  obtain the  coefficient of the flow velocity profile to first order in $g$:
\begin{equation}
u^{(1)}_2 = - \frac {1} {4}.
\label{57}
\end{equation}

The process of solution continues in the same manner for the coefficients of moments of degree $k>2$. We use the relations (\ref{52}) and (\ref{53}) in a first stage to determine the coefficients $\chi^{(1,1,0)}_{\bf k}$  and $\chi^{(1,0,1)}_{\bf k}$  with  $k = 4$.  In the second stage, the obtained results are inserted into (\ref{54}) to evaluate the coefficients of degree $k=3$. The same steps can continue infinitely by following the routes $k=6 \rightarrow k=5$; $k=8 \rightarrow k=7$; \ldots. It is not necessary to evaluate all the coefficients, but some of them are essential to the solution of the problem to the next order in $g$.

\subsection{Second order in $g$ $(s=2)$}

As done in the first-order evaluation, let us take $s=2$ in Eq.\ (\ref{44}) and insert Eqs.\  (\ref{46})--(\ref{49}). We then obtain an equation more complicated than that of the first order.  Of course, the symmetry property $\chi_{\bf k}^{(s,i,j)}=0$ if $i+k_1=\text{odd}$ or $j+k_2=\text{odd}$ must be applied. By equating the coefficients of the same degree in $x$ and $y$  we get the following   set of algebraic equations:
\begin{enumerate}
\item[a)] $i+j=4$
\begin{equation}
 0 = J^{(2,i,j)}_{\bf k},
\label{58}
\end{equation}
\item[b)] $i+j=3$
\begin{equation}
 (i+1) \chi^{(2,i+1,j)}_{k_1+1,k_2,k_3}+(j+1) \chi^{(2,i,j+1)}_{k_1,k_2+1,k_3} = J^{(2,i,j)}_{\bf k},
\label{59}
\end{equation}
\item[c)] $i+j=2$
\beqa
(i+1) \chi^{(2,i+1,j)}_{k_1+1,k_2,k_3}  &+& (j+1) \chi^{(2,i,j+1)}_{k_1,k_2+1,k_3}+2 k_3 u^{(1)}_2 \left[ \chi^{(1,i-1,j)}_{k_1+1,k_2,k_3-1} \right.
\nonumber\\
& + &\left.\chi^{(1,i,j-1)}_{k_1,k_2+1,k_3-1} \right] = J^{(2,i,j)}_{\bf k} ,
\label{60}
\eeqa
\item[d)] $i+j=1$
\beqa
(i+1) \chi^{(2,i+1,j)}_{k_1+1,k_2,k_3}  &+& (j+1) \chi^{(2,i,j+1)}_{k_1,k_2+1,k_3}
+
2 k_3 u^{(1)}_2 \left[ \chi^{(1,i-1,j)}_{k_1+1,k_2,k_3-1}\right.
\nonumber\\
&  
+ &\left.\chi^{(1,i,j-1)}_{k_1,k_2+1,k_3-1} \right]
-
 k_3 \chi^{(1,i,j)}_{k_1,k_2,k_3-1}
 = J^{(2,i,j)}_{\bf k} ,
\label{61}
\eeqa
\item[e)] $i+j=0$
\begin{equation}
\chi^{(2,1,0)}_{k_1+1,k_2,k_3}  + \chi^{(2,0,1)}_{k_1,k_2+1,k_3}- k_3 \chi^{(1,0,0)}_{k_1,k_2,k_3-1}  = J^{(2,0,0)}_{\bf k} .
\label{62}
\end{equation}
\end{enumerate}
In the above equations the convention $\chi^{(2,i,j)}_{\bf k}=0$ if $i<0$ or $j<0$ is implicitly understood.

According to the system (\ref{59})--(\ref{62}),  the coefficients of degree $k+1$ with $i+j=\alpha+1$  determine those of degree $k$ with $i+j=\alpha$. For instance, Eq.\ (\ref{58}) allows one to determine the coefficients with $k=4$ and $i+j=4$. The results obtained are inserted into (\ref{59}) to determine the coefficients of  degree $k=3$ and $i+j=3$. Then we go to Eq.\  (\ref{60}) in order to determine those of degree $k=2$ and $i+j=2$. In general, the process of solution is done according to the following scheme:
\begin{equation}
\{\chi^{(2,i+j=4)}_{k}\} \rightarrow \{\chi^{(2,i+j=3)}_{k-1}\} \rightarrow \{\chi^{(2,i+j=2)}_{k-2}\} \rightarrow \{\chi^{(2,i+j=1)}_{k-3}\} \rightarrow \{\chi^{(2,i+j=0)}_{k-4}\},
\label{63}
\end{equation}
where $\{\chi^{(2,\alpha)}_{k}\}$ denotes the set of coefficients of the same degree, namely $\{\chi^{(2,\alpha)}_{k}\}\equiv \{\chi^{(2,i,j)}_{\bf k}; k_1+k_2+k_3=k, i+j=\alpha\}$.
The expression of $J^{(2,i,j)}_{\bf k}$ is a linear combination of the coefficients $\chi^{(2,i,j)}_{\bf k}$, $T^{(2)}_4$, $T^{(2)}_2$ and $p^{(2)}_2$.
Once a row $k$ of the chain (\ref{63}) is solved, it is necessary to go to the following row $k+2$. This process continues until all the coefficients of the required profiles are completely determined. 
The results for the pressure and temperature are
\beq
p^{(2)}=\frac{3}{10}r^2,\quad T^{(2)}=\frac{34}{175}r^2-\frac{1}{240}r^4.
\label{n4}
\eeq

\subsection{Third and fourth orders in $g$ ($s=3,4$)}
The process to third order in $g$ is practically the same as the previous one.  Since the pressure and the temperature are even functions of $g$,  only the coefficients of $u_z$ appear at this level. The result is
\beq
u^{(3)}=u^{(3)}_2 r^2-\frac{4\,429}{50\,400}r^4-\frac{1}{2\,160} r^6,
\label{n5}
\eeq
where the coefficient $u^{(3)}_2=-6.0806\ldots$ is related  to the ratio $A_4/A_2$.

The  calculation algorithm is essentially  the same for the higher orders in $g$. However, the equations become more and more cumbersome and their solution requires a considerable computational effort. Here we give the final results for the pressure and the temperature to fourth order:
\begin{equation}
p^{(4)}=p^{(4)}_2 r^2+p^{(4)}_4 r^4+ \frac{7}{7\,200} r^6,
\label{66}
\end{equation}
\beq
T^{(4)}=T^{(4)}_2r^2+T^{(4)}_4 r^4-\frac{30\,931}{13\,608\,000} r^6- \frac{23}{691\,200} r^8,
\label{67}
\eeq
where 
 $p^{(4)}_2=-24.160\ldots$, $p^{(4)}_4=-0.062666\ldots$, $T^{(4)}_2=-59.511\ldots$ and $T^{(4)}_4=-0.52856\ldots$.
The high values of $|p^{(4)}_2|$  and $|T^{(4)}_2|$ suggest that the series (\ref{40})--(\ref{39}) are only asymptotic, in agreement with the situation in the case of the BGK model \cite{NOTE:6}.

Along with the hydrodynamic fields, we have obtained all the moments of degrees $k=3$ and $k=4$ to third order in $g$. This allows us to get the pressure tensor and the heat flux. Their expressions in cylindrical coordinates are given in Sec.\ \ref{sec5}.


\begin{thebibliography}{99}
\bibitem{note}
Consider for instance a highly rarefied gas of argon atoms ($m=6.63\times 10^{-26}~\text{kg})$ treated as hard spheres of diameter $3.5~\text{\AA}$. At a number density $n_0=10^{-20}~\text{m}^{-3}$ the mean free path is $\ell_0=0.028~\text{m}$. If the temperature is $T_0=300~\text{K}$, then the value $g=0.1v_0^2/\ell_0$ is equivalent to $g=2.24\times 10^5~\text{m}/\text{s}^2$.
\end{thebibliography}
\end{document}